\documentclass[pra,superscriptaddress,onecolumn,amsmath,amssymb]{revtex4}
\usepackage{amsfonts}
\usepackage{amssymb}
\usepackage{mathtools}
\usepackage{graphicx}
\usepackage[english]{babel}
\usepackage{subfigure}
\usepackage{mathrsfs}
\usepackage{color}
\usepackage[normalem]{ulem}
\usepackage{natbib}
\usepackage{bbold}
\usepackage{placeins}
\usepackage{braket}
\usepackage{soul,xcolor}
\usepackage{epstopdf}
\usepackage{tabu}
\usepackage{array}
\usepackage{bm}
\usepackage{upgreek}
\usepackage{multirow}
\usepackage[utf8]{inputenc}
\usepackage[colorlinks = truelinkcolor = blue, urlcolor  = blue, citecolor = blue, anchorcolor = blue]{hyperref}
\hypersetup{
	colorlinks   = true, 
	urlcolor     = blue, 
	linkcolor    = blue, 
	citecolor   = blue 
}

\makeatletter
\begin{document}


\title{Non-Markovian  channel from the reduced dynamics of coin in quantum walk}
	\author{Javid Naikoo}
\email{naikoo.1@iitj.ac.in}
\affiliation{Indian Institute of Technology Jodhpur, Jodhpur 342011, India}

\author{Subhashish Banerjee}
\email{subhashish@iitj.ac.in}
\affiliation{Indian Institute of Technology Jodhpur, Jodhpur 342011, India}

\author{C. M. Chandrashekar}
\email{chandru@imsc.res.in }
\affiliation{The  Institute  of  Mathematical  Sciences,  C.  I.  T.  Campus,  Taramani,  Chennai  600113,  India}
\affiliation{Homi  Bhabha  National  Institute,  Training  School  Complex,  Anushakti  Nagar,  Mumbai  400094,  India}



\begin{abstract}
	\noindent 
	The quantum channels with memory, known as non-Markovian channels,  are of crucial importance for a  realistic description of a variety of physical systems, and  pave ways for new methods of decoherence control by manipulating the properties of environment such as its frequency spectrum. In this work, the reduced dynamics of coin in a discrete-time quantum walk is characterized as a non-Markovian  quantum channel. A general formalism is sketched to extract the Kraus operators for a $t$-step quantum walk. Non-Markovianity, in the sense of P-indivisibility of the reduced coin dynamics, is inferred from the non-monotonous behavior of  distinguishably of two orthogonal states subjected to it. Further, we study various quantum information theoretic quantities of  a qubit under the action of this channel,  putting in perspective, the role such channels can play in various quantum information processing tasks.
\end{abstract}


\maketitle

\section{Introduction}

The study of open quantum systems with memory has  attracted lot of attention over last few years, since such systems describe a plethora of  physical phenomena  and also provide new ways to control various quantum features by engineering the system-environment interactions \cite{BP,sbbook}. Several investigations on the  role of structured environments and non-Markovianity in entanglement generation \cite{HuelgaNMent},  quantum teleportation \cite{laine2014nonlocal}, 	key distribution \cite{NMkeydist},  quantum metrology \cite{NMmetrology}, quantum biology \cite{NMqbiology}, have suggested the advantage of  non-Markovian quantum channels over Markovian ones. 

  Quantum walks (QWs) was conceived as  a generalization of classical random walks with an anticipation of its potential in modeling the dynamics particle in quantum realm \,\cite{Riazanov,Feynman,Parthasarathy,Meyer,mayer2016, AharonovQRW}.  They describe the coherent evolution of a quantum particle, coin space coupled to the position space which in principle can be treated as an external environment.  One-dimensional QWs involve a walker free to move in either direction along a straight line such that the direction for each step is decided  by the outcome of a coin. However, it differs from its classical counterpart in  the sense that the probability distribution of the quantum particle spreads quadratically faster in position space than the classical random walk due to interference.  This feature makes QWs ideal candidate for development of quantum algorithms such as  quantum search algorithms \cite{shenvi2003quantum,krovi2006hitting}. Ability to engineering the dynamics of the QWs  has also allowed us to simulate  and study quantum correlations \cite{mid,entanglement,discord}, quantum to classical transition \cite{sbQWnoise,SBQWsymmetry}, memory effects and disorder \cite{kumar2018enhanced}, relativistic quantum effects \cite{sbchandrudirac} and quantum games \cite{parrondo}.  Experimental implementation of QWs has been realized in various  physical systems viz., in cold atoms \cite{ExpQWwaveguide, ExpQWoptTrpAtoms},  photonic systems \cite{peruzzo2010quantum,schreiber2010photons,broome2010discrete,tamura2020quantum,kitagawa2012observation,preiss2015strongly,cote2006quantum}. Recent studies have reported the circuit based implementation of QW  \cite{ryan2005experimental,qiang2016efficient,alderete2020quantum}. A scheme for  implementing  QW in Bose-Einstein condensates was presented in \cite{chandrashekar2006implementing} and was  recently implemented in momentum space \cite{ExpBEcondensate}.  Possible applications of QWs   in understanding the dynamics in biological systems  have been reported in various works \cite{hoyer2010limits,mohseni2008environment,rebentrost2009environment}.

 The QW can be discrete or continuous in time, accordingly known as Discrete  Time Quantum Walk (DTQW) and  Continuous Time Quantum Walk (CTQW). In this work, we confine ourselves to the former case. The DTQW was studied from the perspective of various facets of non-Markovian evolution, such as the disambiguation of contributions  to  non-Markovian   backflow as well as the transition from quantum to classical random walks \cite{PradeepOSID}. The non-Markovian nature of coin dynamics in DTQW can be brought out by tracing over the position space \cite{hinarejos}. Henceforth, we will coin the term quantum walk noise (QWN) to describe the reduced dynamics on the coin space. In this work, we quantify this by developing the Kraus operators for the QWN, thereby characterizing the QW channel. The QWN was studied \cite{PradeepOSID} in conjunction with an RTN \cite{daffner,rice} noise. The P-indivisibility \cite{RHP, BLP,SBSrik} of the QWN as well as the RTN suggested that the intermediate map of the full evolution could be not completely positive (NCP). Also, non-monotonic behavior under trace distance was indicated. This called for a careful consideration of the application of such non-Markovian noise channels to the DTQW protocol. A suggestion offered in \cite{PradeepOSID} was that in contrast to the conventional application of the (Markovian) noise channel \cite{sbQWnoise,SBQWsymmetry} in the form of appropriate Kraus operators \cite{sbbook},  after each application of the walk operation, in the present non-Markovian scenario, the Kraus operators  are applied once after $t$ QW steps. This notion was implemented numerically.  Here, making use of the developed Kraus operators of the QW channel, we quantify this notion. This, thus also serves the purpose of highlighting the implementation of non-Markovian noise channels to various QW protocols. We further characterize the QW channel by studying various information theoretic processes on it. Specifically, the interplay of purity of qubit state with the channel parameter as well as the state parameter is investigated. Further, the Holevo quantity, which characterizes the information about an input state that can be retrieved from the output of the channel, is studied.

The paper is organized as follows: In Sec. (\ref{sec:dynamics}), the reduced coin dynamics is studied, sketching the formalism to extract the Kraus operators for a $t$-step walk. Section (\ref*{sec:properties}) is devoted to a detailed investigation of  various properties of QW channel, such as its non-Markovian nature in the sense of P-indivisibility in Sec.  (\ref{subsec:NM}), the purity of states subjected to this  channel in Sec.  (\ref{subsec:purity}), and the Holevo quantity in Sec. (\ref{subsec:Holevo}). Conclusion of this work is presented  in Sec. (\ref{sec:conclusion}).


\section{Reduced  Dynamics of Coin}\label{sec:dynamics}


			  Let the initial state of  coin and  walker be $\ket{\psi_c}$ and $\ket{\psi_p}$, respectively. The unitary operator $\hat{W} = \hat{S} (\hat{C} \otimes \mathbb{1})$, where $\hat{S}$ and $\hat{C}$ are the shift and coin operators, respectively, governs the time evolution of the combined state $\ket{\psi_c} \otimes \ket{\psi_p}$. The state after $t$ steps is given by \cite{chandrashekar2010discrete}
			\begin{equation}\label{eq:psit}
			\ket{\psi(t)}  =  \hat{W}^t (\ket{\psi_c} \otimes \ket{\psi_p}),\quad {\rm or} \quad \rho(t) = \hat{W}^t (\rho_c \otimes \rho_p) (W^t)^\dagger.
			\end{equation}
			Here, $\rho(t) = | \psi(t) \rangle \langle \psi(t) |$, $\rho_p = |\psi_p \rangle \langle \psi_p |$, and $\rho_c = | \psi_c \rangle \langle \psi_c|$ are the corresponding density matrices.  Further, the coin and shift operators are given by
			\begin{align}\label{eq:CS}
			\hat{C} &= \begin{pmatrix}
			\cos\theta    &   -i \sin\theta   \\
			-i \sin\theta  &   \cos\theta
			\end{pmatrix}, ~~ {\rm and}	~~~~	 
			\hat{S} =  \ket{\uparrow}\bra{\uparrow} \otimes \hat{S}_L + \ket{\downarrow}\bra{\downarrow} \otimes \hat{S}_R.
			\end{align}           
			The operator  $\hat{S}_L= \sum\limits_{x \in \mathbb{Z}} \ket{x-1}\bra{x}$, and   $\hat{S}_R= \sum\limits_{x \in \mathbb{Z}} \ket{x+1}\bra{x}$, are the left and right shift operators, respectively.  The total unitary operator for $t$ steps becomes 
			\begin{align}\label{eq:Wt}
			W^t &= \big[\hat{S}  (\hat{C} \otimes \mathbb{1}) \big]^t \nonumber \\
			       &=\Big[ \Big( \ket{\uparrow}\bra{\uparrow} \otimes \hat{S}_L + \ket{\downarrow}\bra{\downarrow} \otimes \hat{S}_R \Big) \Big(\hat{C} \otimes \mathbb{1} \Big) \Big]^t \nonumber \\
			       &= \Big[ \ket{\uparrow}\bra{\uparrow} \hat{C} \otimes \hat{S}_L + \ket{\downarrow}\bra{\downarrow} \hat{C} \otimes \hat{S}_R  \Big]^t \nonumber \\
			       &=  \Big[  \hat{C}_{\uparrow} \otimes \hat{S}_L + \hat{C}_{\downarrow} \otimes \hat{S}_R  \Big]^t = \Big[\hat{P} + \hat{Q}\Big]^t. \nonumber \\
			\end{align}
			Here, $\hat{P} =  \hat{C}_{\uparrow} \otimes \hat{S}_L$,  $\hat{Q} = \hat{C}_{\downarrow} \otimes \hat{S}_R $,  $  \hat{C}_{\uparrow}  = \ket{\uparrow}\bra{\uparrow} \hat{C} $, and $ \hat{C}_{\downarrow} =  \ket{\downarrow}\bra{\downarrow} \hat{C}$.  The right hand side can be  simplified using the \textit{binomial} expansion \cite{wyss2017two}
			\begin{equation}\label{eq:Binomial}
			(\hat{P} + \hat{Q})^t = \sum\limits_{k=0}^{t} \binom{t}{k} \hat{P}^k \hat{Q}^{t-k} +  \sum\limits_{k=0}^{t} \binom{t}{k} \hat{D}_k(\hat{Q}, \hat{P}) \hat{Q}^{t-k}.
			\end{equation}
			The second term arises due  to the non-commutative nature of $\hat{P}$ and $\hat{Q}$, and can be simplified using the recurrence relation
			\begin{equation}\label{eq:Dkplus1}
			\hat{D}_{k+1}(\hat{Q}, \hat{P}) = [\hat{Q}, \hat{P}^k] + \hat{P} \hat{D}_k(\hat{Q}, \hat{P}) + [\hat{Q}, \hat{D}_k(\hat{Q}, \hat{P})], \quad {\rm with} \quad \hat{D}_0 (\hat{Q}, \hat{P}) = 0.
			\end{equation}
			Thus, the quantity $	\hat{D}_{k+1}(\hat{Q}, \hat{P}) $ vanishes   if  $[\hat{Q}, \hat{P}] = 0$.  From the definition of $\hat{P}$ and $\hat{Q}$, it follows
			 \begin{equation}\label{eq:QPPQ}
			[\hat{Q}, \hat{P}] =     \hat{Q} \hat{P} - \hat{P} \hat{Q} =  \hat{C}_{\downarrow} \hat{C}_{\uparrow} \otimes \mathbb{1} -  \hat{C}_{\uparrow}  \hat{C}_{\downarrow}  \otimes \mathbb{1}.
			 \end{equation}
			 Using the definition of $ \hat{C}_{\downarrow (\downarrow)}$, it  follows that $	[\hat{Q}, \hat{P}] = \begin{pmatrix} -\sin^2\theta & -i \sin\theta \cos\theta \\  i \sin\theta \cos\theta & \sin^2\theta \end{pmatrix}$, and is a zero matrix only for $\theta = 0$, $\pi$, and  $2\pi$, which correspond to the coin operator being  identity.
			
			Further simplification of the first term  in Eq. (\ref{eq:Binomial}) reads
			\begin{align}
             		\sum\limits_{k=0}^{t} \binom{t}{k} \hat{P}^k \hat{Q}^{t-k}	  &= \sum\limits_{k} \binom{t}{k} ( \hat{C}_{\uparrow} \otimes \hat{S}_L)^{t-k} ( \hat{C}_{\downarrow} \otimes \hat{S}_R )^k  = \sum\limits_{k} \binom{t}{k}  \hat{C}_{\uparrow}^{t-k}  \hat{C}_{\downarrow}^{k} \otimes \hat{S}_L^{t-k} \hat{S}_R ^k.
			\end{align}

			 For a walk of   $t$-steps,  symmetric about $x=0$,  the number of values position can take is $ 2 t +1$.   Let the initial state of coin and walker be $\ket{\psi_c} = a \ket{\uparrow} + b\ket{\downarrow}$ (with $|a|^2 + |b|^2 = 1$) and $\ket{\psi_p} = \ket{x=0}$, respectively. The possible position states are $\ket{x=-t}, \dots, \ket{x=t} $. We represent these states in computational basis as $(1~0~0 \dots )^T$, $\dots$, $(0~0 \dots 1)^T$, respectively.

\bigskip
	
			 With this setting, we trace over the position degrees of freedom, using notation $| x = \mu  \rangle = | x_\mu \rangle$, and obtain
			\begin{equation}
			\rho_c(t) = \sum\limits_{\mu =-t}^{t} \langle x_\mu | \hat{W}^t (\rho_c \otimes  |\psi_p \rangle \langle \psi_p | ) (\hat{W}^t)^\dagger | x_\mu \rangle =  \sum\limits_{\mu =-t}^{t} K_\mu \rho_c K_\mu^\dagger.
			\end{equation}     
			The Kraus operators are identified as , with $\mu = -t, \dots,  t$.
			\begin{align}
              K_\mu &=  \langle x_\mu |  \hat{W}^t | \psi_p \rangle = \langle x_\mu |	(\hat{P} + \hat{Q})^t |\psi_p \rangle \nonumber \\
                   &= \sum\limits_{k=0}^{t} \binom{t}{k}  \langle x_\mu | \hat{P}^k \hat{Q}^{t-k} | \psi_p \rangle +   \sum\limits_{k=0}^{t} \binom{t}{k}  \langle x_\mu | \hat{D}_k(\hat{Q}, \hat{P}) \hat{Q}^{t-k} |\psi_p \rangle.
			\end{align}
		   In order to simplify the first term, we  assume $\ket{\psi_p} = \ket{0}$, i.e., the walker starts at $x=0$, such that
			\begin{align}
               \sum\limits_{k=0}^{t} \binom{t}{k}  \langle x_\mu | \hat{P}^k \hat{Q}^{t-k} | 0 \rangle &=   \sum\limits_{k=0}^{t} \binom{t}{k}  \langle x_\mu | \Big( \hat{C}_{\uparrow} \otimes \hat{S}_L\Big)^k \Big( \hat{C}_{\downarrow} \otimes \hat{S}_R \Big)^{t-k}  | 0 \rangle \nonumber \\
               &= \sum\limits_{k=0}^{t} \binom{t}{k} \hat{C}_{\uparrow}^k \hat{C}_{\downarrow}^{t-k}  \langle x_\mu | \hat{S}_L^k   \hat{S}_R^{t-k} | 0 \rangle \nonumber \\
               &=  \sum\limits_{k=0}^{t} \binom{t}{k} \hat{C}_{\uparrow}^k \hat{C}_{\downarrow}^{t-k}  \delta_{\mu+k, t- k}\nonumber \\
               &= \frac{t!}{(\frac{t-\mu}{2})! (\frac{t+\mu}{2})!} \hat{C}_{\uparrow}^{\frac{t-\mu}{2}} \hat{C}_{\downarrow}^{\frac{t+\mu}{2}}.
			\end{align}
			Use has been made of $\langle x_\mu | \hat{S}_L^k   \hat{S}_R^{t-k} | 0 \rangle =  \delta_{\mu+k, t- k}$, see the Appendix. The constraints $k = (t- \mu )/2$ and $k \in  \{0,1,2, \dots \}$ demand that $\mu$ and $t$ have same parity, i.e., for  $t$  even (odd),  $\mu$  is even (odd).
			
			\bigskip

			For a one step walk,  $t=1$, implies  $\mu = -1, 1$. From Eq. (\ref{eq:Dkplus1}) $D_1(\hat{P}, \hat{Q}) = 0$, we have  $K_\mu =  \frac{t!}{(\frac{t-\mu}{2})! (\frac{t+\mu}{2})!} \hat{C}_{\uparrow}^{\frac{t-\mu}{2}} \hat{C}_{\downarrow}^{\frac{t+\mu}{2}} $, leading to
			
			\begin{equation}
			K_{-1} = \begin{pmatrix}
			0                        &       0\\
			-i \sin\theta  & \cos\theta           
			\end{pmatrix},\quad   K_1 = \begin{pmatrix}
			\cos\theta  & -i \sin\theta  \\
			0                        &    0         
			\end{pmatrix}.
			\end{equation}         
			These operators satisfy the completeness condition $K_{-1}^\dagger K_{-1} +  K_1^\dagger K_1 = \mathbb{1}$.   Table (\ref{tab:symQW}) lists the Kraus operators for the reduced coin dynamics for  a few steps of symmetric QW.  One infers that,
			\begin{enumerate}
				\item     $K_{-t} = \mathcal{M}[K_{t}]$,  where $ \mathcal{M}[K_{t}]$ is the \textit{minor} of the matrix $K_t$.
				\item  For coin parameter $\theta = \pi/2$, $K_{\pm 2n} = 0$, $n=1,2,3\dots$, and $K_0 = \pm \mathbb{1}$, with $ \mathbb{1}$ being the identity matrix.
			\end{enumerate}
			\bigskip

			\begin{figure} 
				\centering
				\begin{tabular}{cc}
					\includegraphics[width=80mm]{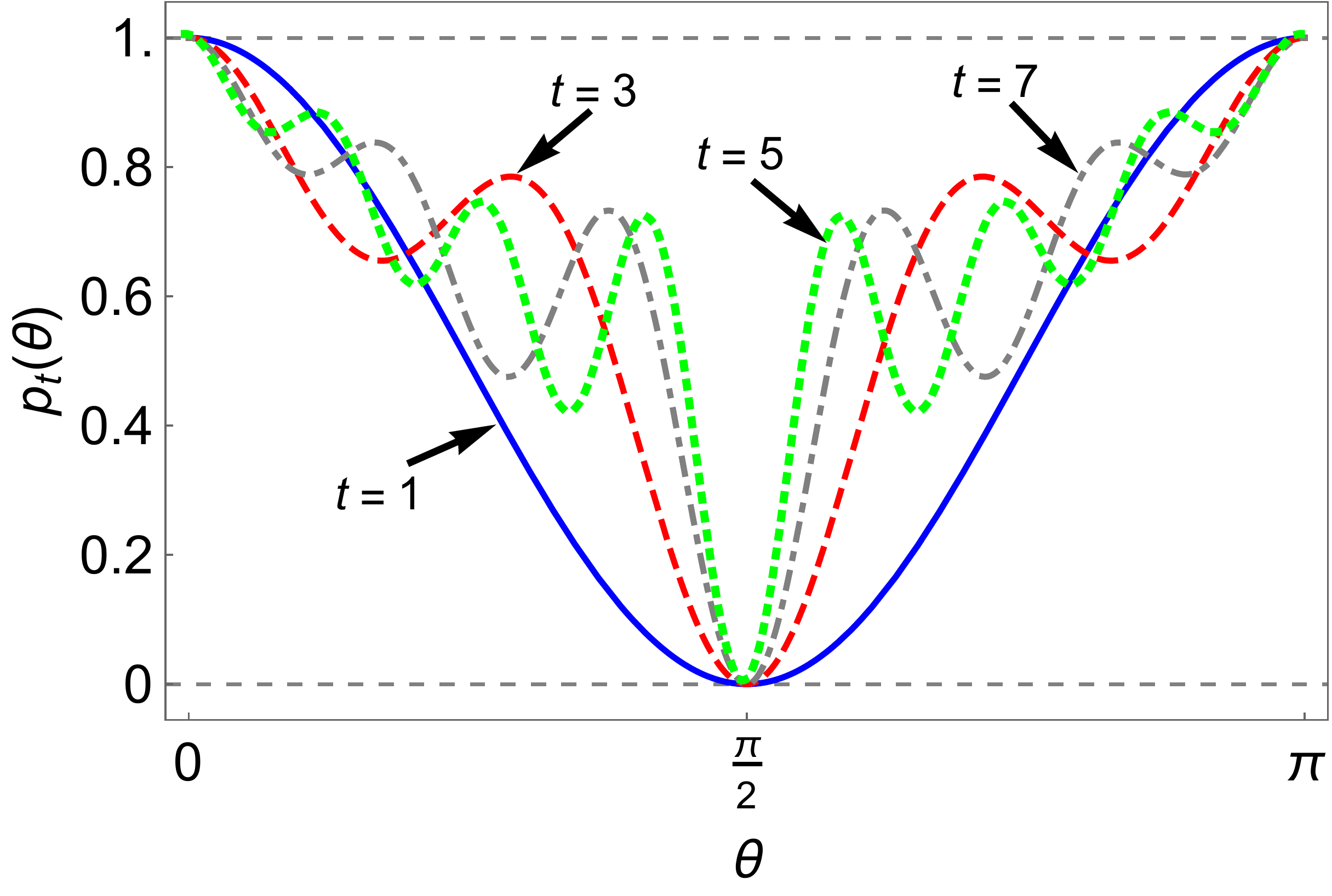} & \includegraphics[width=80mm]{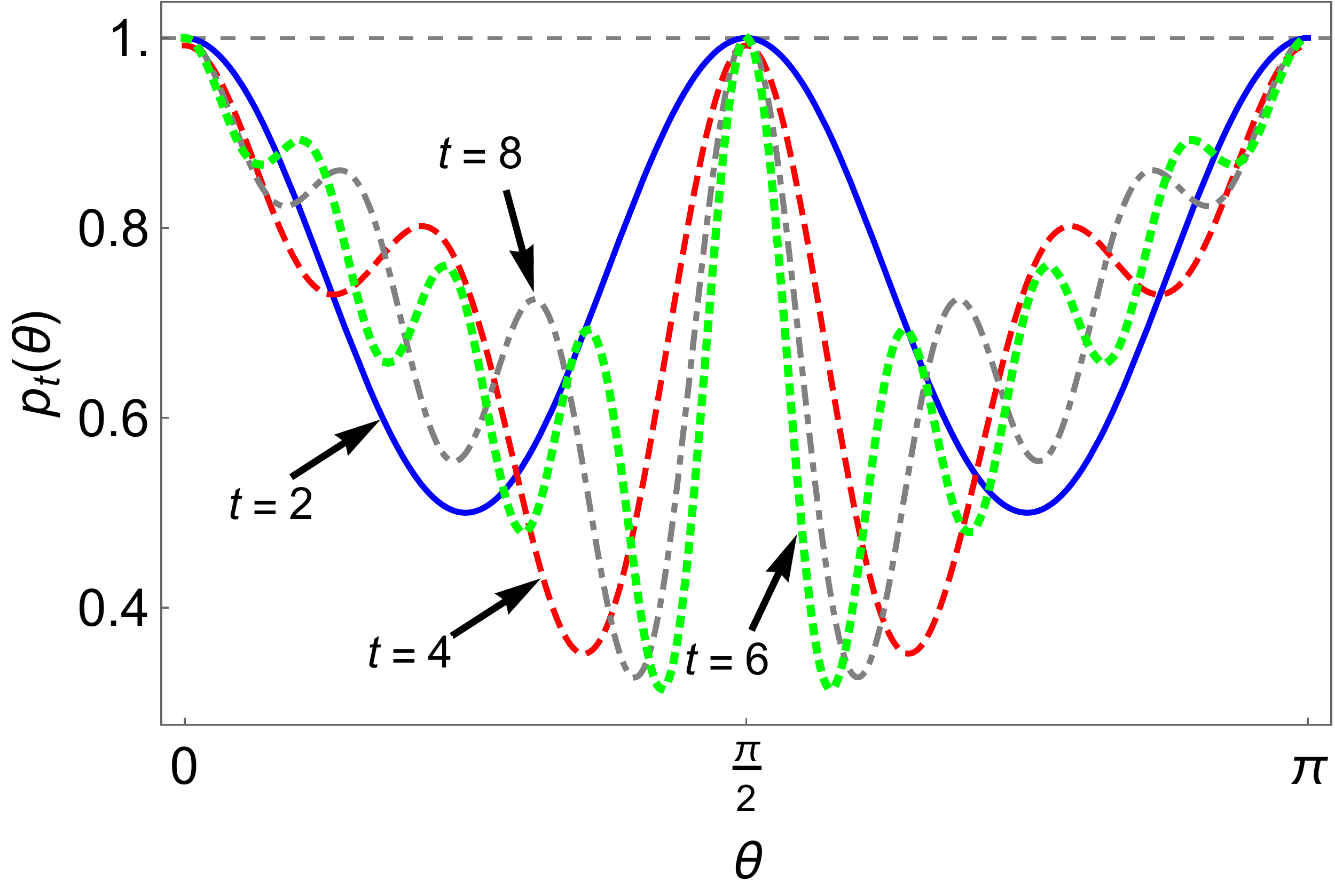}
					\tabularnewline
					(a)  & (b) \tabularnewline
					\includegraphics[width=80mm]{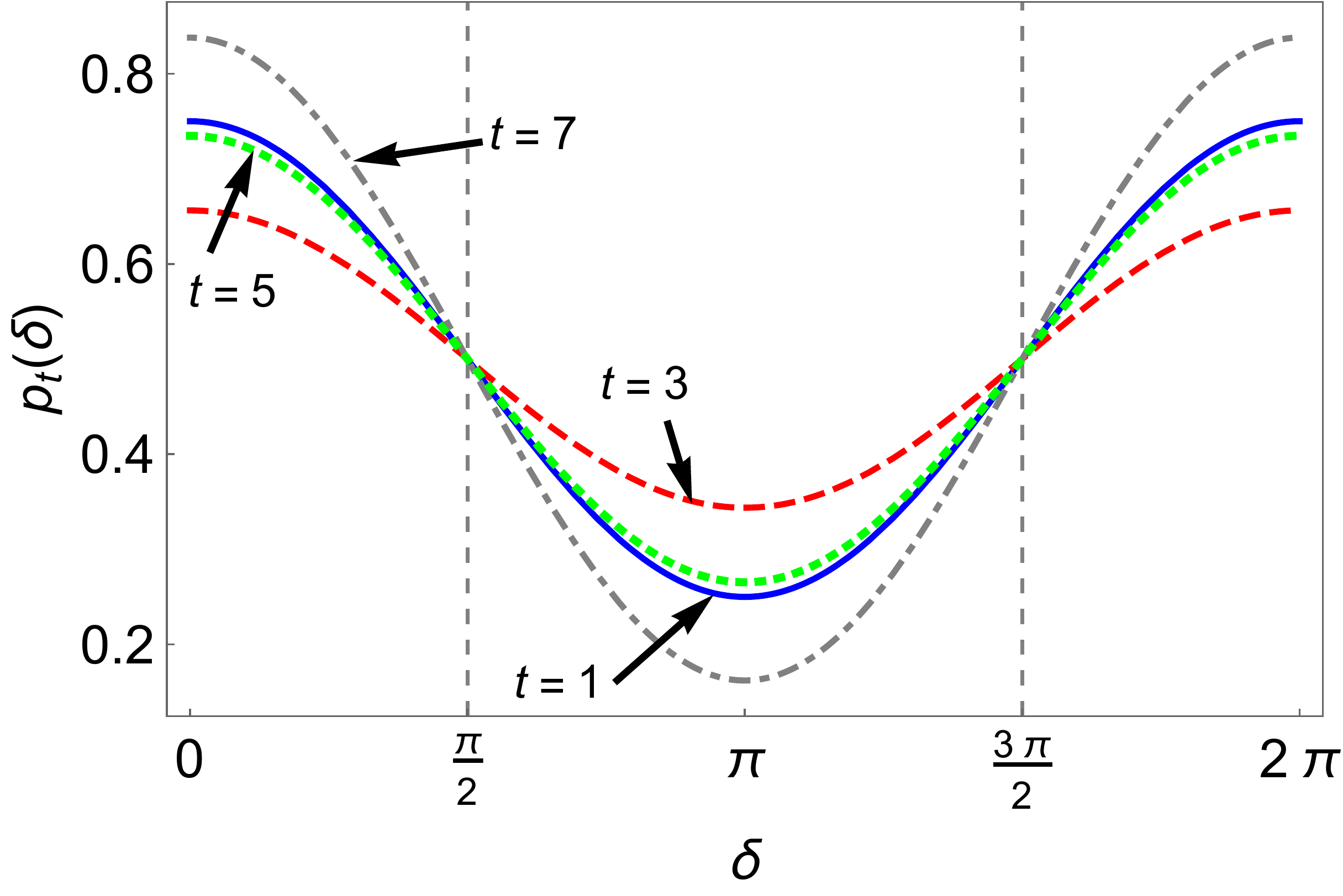} & \includegraphics[width=80mm]{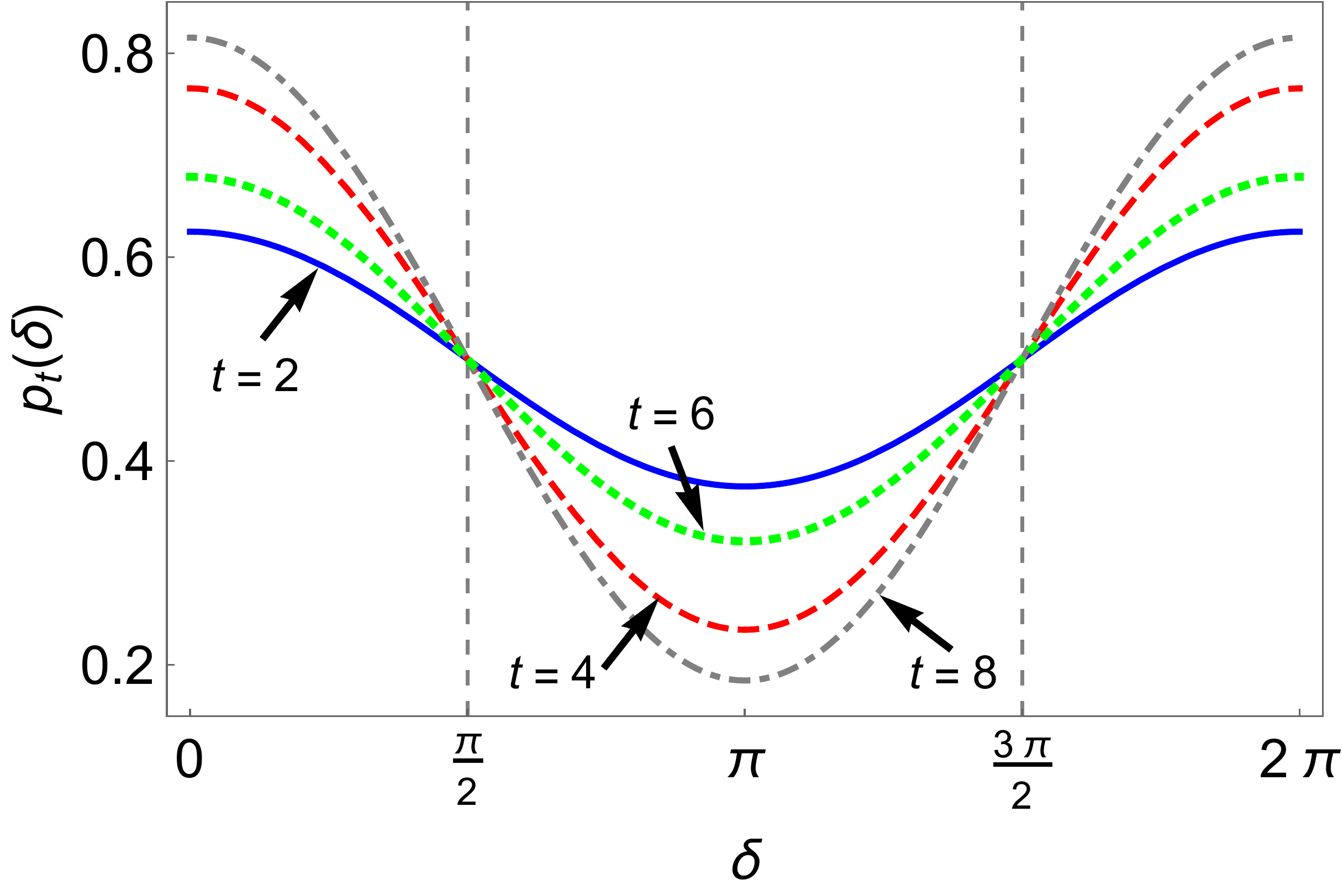}
					\tabularnewline
					(c)  & (d) \tabularnewline
				\end{tabular}
				\caption{(Color online) Depicting probability $p_t$  (see Eq. (\ref{eq:coinmap})) of obtaining $\ket{0}$ in an $t$-step QW with respect to the coin parameter $\theta$  (a)-(b) with initial state  $\ket{\psi_c} =\ket{0}$, and with respect to the state parameter $\delta$ in (c)-(d) with initial state $\ket{\psi_c} = \cos(\delta/2) \ket{0} + \sin(\delta/2) \ket{1}$, and  coin parameter $\theta = \pi/6$.}
				\label{fig:prob}
			\end{figure}

			  The  Kraus operators $K_{t}$ constitute a map $\mathcal{F}$ connecting the  input state $\rho_c(0) $ to output $\rho_c(t)$. Let $\rho_c(0) = \ket{\psi_c(0)}\bra{\psi_c(0)}$ with $\ket{\psi_c(0)} = a \ket{\uparrow} + b \ket{\downarrow}$, we have
			\begin{equation}\label{eq:coinmap}
			\rho_c(0) = \begin{pmatrix}
			|a|^2  &  a b^*\\
			a^* b  & |b|^2      
			\end{pmatrix}  \rightarrow \rho_c(t) =[ \mathcal{F}]_{t=n} \rho_c(0) = \sum\limits_{\mu = -t}^{t}  K_{\mu} \rho_c(0) K_{\mu}^\dagger = \begin{pmatrix}
			p_t(\theta)   &  q_t(\theta)\\
			q_t^*(\theta)  &  1 - p_t(\theta)
			\end{pmatrix}.
			\end{equation}
			Here, $p_{t}(\theta)$ is the probability of obtaining $\ket{\uparrow}$  in an $t$-step walk. The form of $p_t(\theta)$ and $q_t(\theta)$ for some steps is given below
			\begin{equation}\label{eq:probs}
			\left.
			\begin{split}
			p_1(\theta) &= | a \cos\theta - b \sin\theta|^2 = \frac{1}{2}\big[ 1 + (|a|^2 - |b|^2) \cos(2\theta) + i (a b^* - a^* b)\sin(2 \theta) \big]\\
			p_2(\theta) &= \frac{1}{4}\big[ 1 + 2 |a|^2 + (|a|^2 - |b|^2) \cos(4\theta) + i (a b^* - a^* b)\sin(4 \theta)\big]\\
			p_3(\theta) &= \frac{1}{16} \big[ 6 + 4 |a|^2 + 5 (|a|^2 - |b|^2) \cos(2\theta) -2 (|a|^2 - |b|^2) \cos(4\theta) + 3 (|a|^2 - |b|^2) \cos(6\theta)  \\&+ 3i (ab^* - a^* b)\sin(2\theta) - 2i (ab^* - a^* b)\sin(4\theta) + 3i (ab^* - a^* b)\sin(6\theta)  \big]
			\end{split}
			\right\}
			\end{equation}
			and 
			\begin{equation}
			\left.
			\begin{split}
			q_1(\theta) &=0\\
			q_2(\theta) &= \sin^2\theta \big[ ab^* \cos^2\theta + a^*b \sin^2\theta + i (|a|^2 - |b|^2)\sin\theta \cos\theta  \big]\\
			q_3(\theta) &= \cos\theta \sin^2\theta \big[(a^*b + ab^*) \cos\theta + (ab^* - a^* b) \cos(3\theta) + i(|a|^2 - |b|^2)\sin(3\theta)\big]
			\end{split}
			\right\}
			\end{equation}
			
			The probabilities $p_t(\theta)$ are plotted in  Fig.  \ref{fig:prob} (a)-(b) when $\ket{\uparrow} = \ket{0}$, with respect to the coin parameter $\theta$. The asymmetric behavior of the probabilities,  with respect to even and odd number of steps, is observed at  $\theta = \pi/2$, where probabilities converge to one (zero) for even (odd) number of steps. The value of the coin parameter $\theta = \pi/2$ corresponds to the coin operator ( Eq. (\ref{eq:CS}) )  $\hat{C} = -i  \sigma_x$, where $\sigma_x$ is the Pauli operator.			
				\begin{table*}[htp]
				\centering
				\caption{\ttfamily Kraus operators for the reduced coin dynamics for some steps of symmetric QW. Here, $\theta$ is the coin parameter defined in Eq. (\ref{eq:CS}).
				}
				\begin{tabular}{ |p{1cm}|p{10cm}|}
					\hline
					Steps       & Kraus operators \\
					\hline
					1               &  \parbox{3cm}{\begin{align*}
						                                          	K_{-1} = \begin{pmatrix}
						                                          0                        &       0\\
						                                          -i \sin\theta  & \cos\theta           
						                                          \end{pmatrix}\quad   K_1 = \begin{pmatrix}
						                                          \cos\theta  & -i \sin\theta  \\
						                                          0                        &    0         
						                                          \end{pmatrix}
						                                          \end{align*}}     \\
					\hline
					2               &  	\parbox{3cm}{\begin{align*}
							                                       K_{-2} &= \begin{pmatrix}   0  & 0 \\ -i \cos\theta \sin\theta & \cos^2\theta \end{pmatrix} \\
						                                             K_{0} &= \begin{pmatrix}  - \sin^2\theta & -i \sin\theta \cos\theta  \\  -i \sin\theta \cos\theta  &  -\sin^2\theta  \end{pmatrix}\\
						                                             K_{2} &= \begin{pmatrix}   \cos^2\theta & -i \sin\theta \cos\theta  \\ 0 &  0 \end{pmatrix}
						                                     \end{align*}}                                                                                                                                \\
					\hline
					3               &    	\parbox{3cm}{\begin{align*}
						                                            	K_{-3} &= \begin{pmatrix}   0  & 0 \\ -i \cos^2\theta \sin\theta & \cos^3\theta \end{pmatrix}\\
						                                            K_{-1} &= \begin{pmatrix}  - \cos\theta \sin^2\theta  & -i \cos^2\theta \sin\theta \\  -i \cos^2\theta \sin\theta + i \sin^3\theta  & -2 \cos\theta \sin^2\theta  \end{pmatrix}\\
						                                            K_{1} &= \begin{pmatrix}  - 2\cos\theta \sin^2 \theta  & -i \cos^2\theta \sin\theta + i\sin^3\theta \\  -i \cos^2\theta \sin\theta   & - \cos\theta \sin^2\theta  \end{pmatrix}\\
						                                            K_{3} &= \begin{pmatrix}  \cos^3\theta   & -i \cos^2\theta \sin\theta \\  0   & 0  \end{pmatrix}
					                                   	          \end{align*}} 
				                                   	            \\
					\hline
					4                &    \parbox{3cm}{\begin{align*}
						                                          	K_{-4} &= \begin{pmatrix}   0  & 0 \\ -i  \cos^3\theta \sin\theta & \cos^4\theta \end{pmatrix}\\
						                                          K_{-2} &= \begin{pmatrix}  - \cos^2\theta \sin^2\theta  & -i \cos^3\theta \sin\theta \\  -i \cos^3\theta \sin\theta + 2 i \cos\theta \sin^3\theta  & -3 \cos^2\theta \sin^2\theta  \end{pmatrix}\\
						                                          K_0 &= \begin{pmatrix}  - 2 \cos^2\theta \sin^2\theta + \sin^4 \theta   & -i \cos^3\theta \sin\theta + 2i \cos\theta \sin^3\theta \\  -i \cos^3\theta \sin\theta + 2 i \cos\theta \sin^3\theta    & - 2\cos^2\theta \sin^2\theta + \sin^4\theta  \end{pmatrix}\\
						                                          K_{2} &= \begin{pmatrix}  - 3 \cos^2\theta \sin^2\theta  & -i \cos^3\theta \sin\theta + 2i \cos\theta \sin^3\theta \\  -i \cos^3\theta \sin\theta    & -\cos^2\theta \sin^2\theta  \end{pmatrix}\\
						                                          K_4 &= \begin{pmatrix}   \cos^4\theta  & -i \cos^3\theta \sin\theta \\ 0 & 0 \end{pmatrix}
						                                         \end{align*}}   \\
					\hline
				\end{tabular}\label{tab:symQW}
			\end{table*}		
		 
		 There are other formulation of QW, like the \textit{split-step} QW, where one breaks each step of walk into two half-step evolutions described by the unitary  $\hat{W}_{ss} = \hat{S}_{+} (\hat{C} \otimes \mathbb{1}) \hat{S}_{-} (\hat{C} \otimes \mathbb{1}) = \hat{W}^2$ \cite{singh2020universal}.  		Here, $\hat{W}$ is the unitary operator for the standard QW, defined in Eq. (\ref{eq:psit}).     The Kraus operators for some steps of the split-step QW are given in Table (\ref{tab:symSSQW}).

	  \section{Some properties of the QW channel}\label{sec:properties}
	  In this section, we characterize the non-Markovian QW channel comprising the reduced coin dynamics. We also study some quantum information theoretic quantities on it.
	  
	  \subsection{Non-Markovian dynamics}\label{subsec:NM}
	  Non-Markovianity is a multifaceted phenomenon. Here, we restrict ourselves to the P-indivisibility form of non-Markovianity, that can be probed by using some state distinguishablity measure, such as   trace distance, denoted by $\mathcal{D}$. Trace distance of  states $\rho$ and $\sigma$ is defined as  $\mathcal{D} (\rho,\sigma) = \frac{1}{2} \sum_i |\lambda_i|$, where $\lambda_i$ are the eigenvalues of  matrix $\rho - \sigma$. A departure from the monotonic behavior of $\mathcal{D} (\mathcal{A}(\rho),\mathcal{A}(\sigma)) $ implies  P-indivisibility of the map $\mathcal{A}$, and hence non-Markovian dynamics.  Consider  two orthogonal  states $\rho_0(t=0) = |0\rangle \langle 0|$ and $\rho_1(t=0) = | 1 \rangle \langle 1|$, subjected to the  QW channel for specific number of steps. For a one step walk, we have
	\begin{equation}
	\mathcal{D}(\rho_0(n=1), \rho_1(n=1)) = \frac{1}{2}\sum_i |\lambda_i| = |\cos(2\theta)|.
	\end{equation}
	Here, $\lambda_i$ are the eigenvalues of $\rho_0(n=1) - \rho_1(n=1)$ and
	\begin{equation}
	\rho_0 (n=1) = \sum\limits_{\mu=1,3} K_\mu \rho_0 K_\mu^\dagger\qquad ~{\rm and }~~~ \rho_1 (n=1) = \sum\limits_{\mu=1,3} K_\mu \rho_1 K_\mu^\dagger.
	\end{equation}
	
\begin{figure}
			\begin{tabular}{ccc}
			\includegraphics[width=50mm]{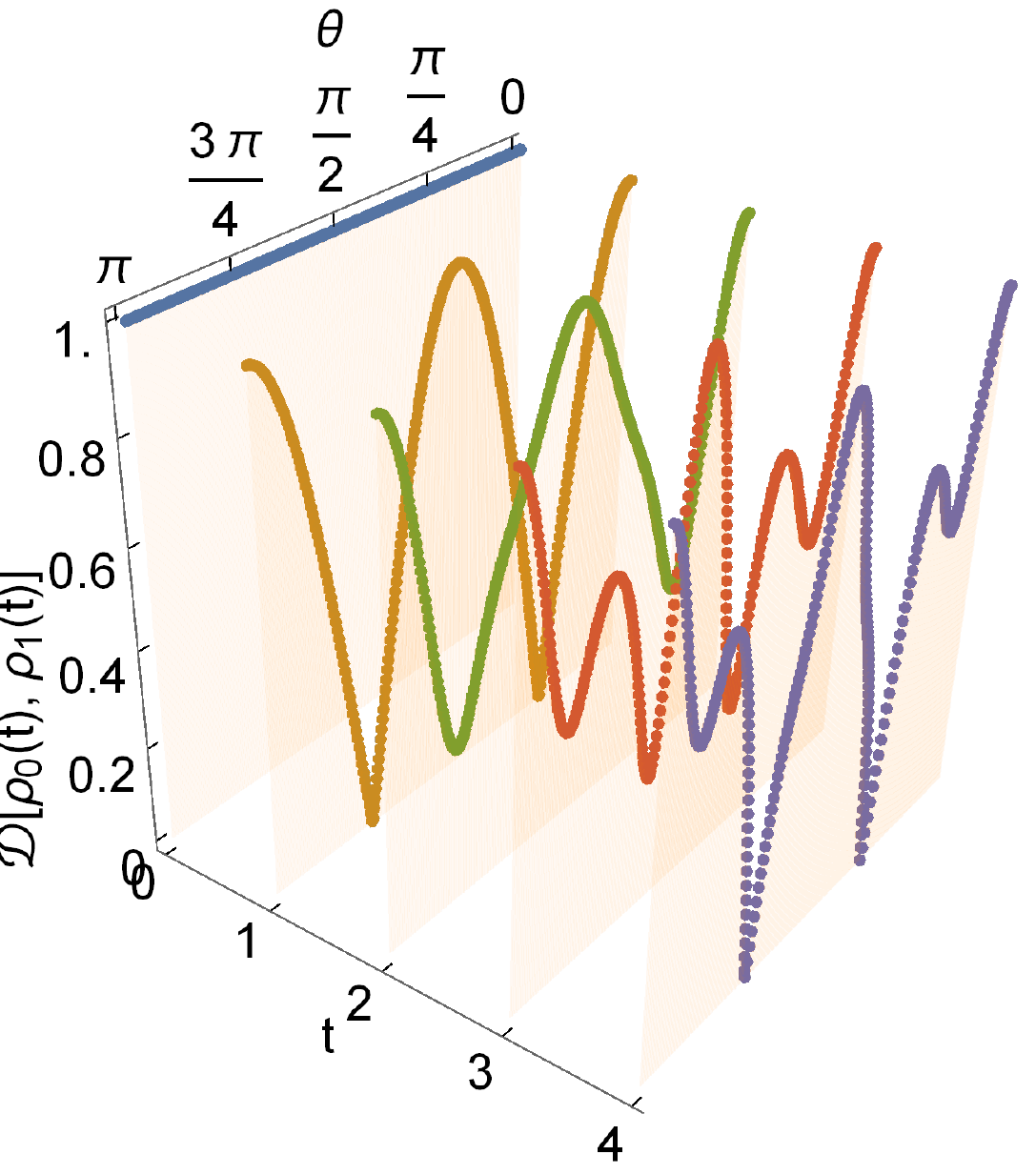} & 	\includegraphics[width=50mm]{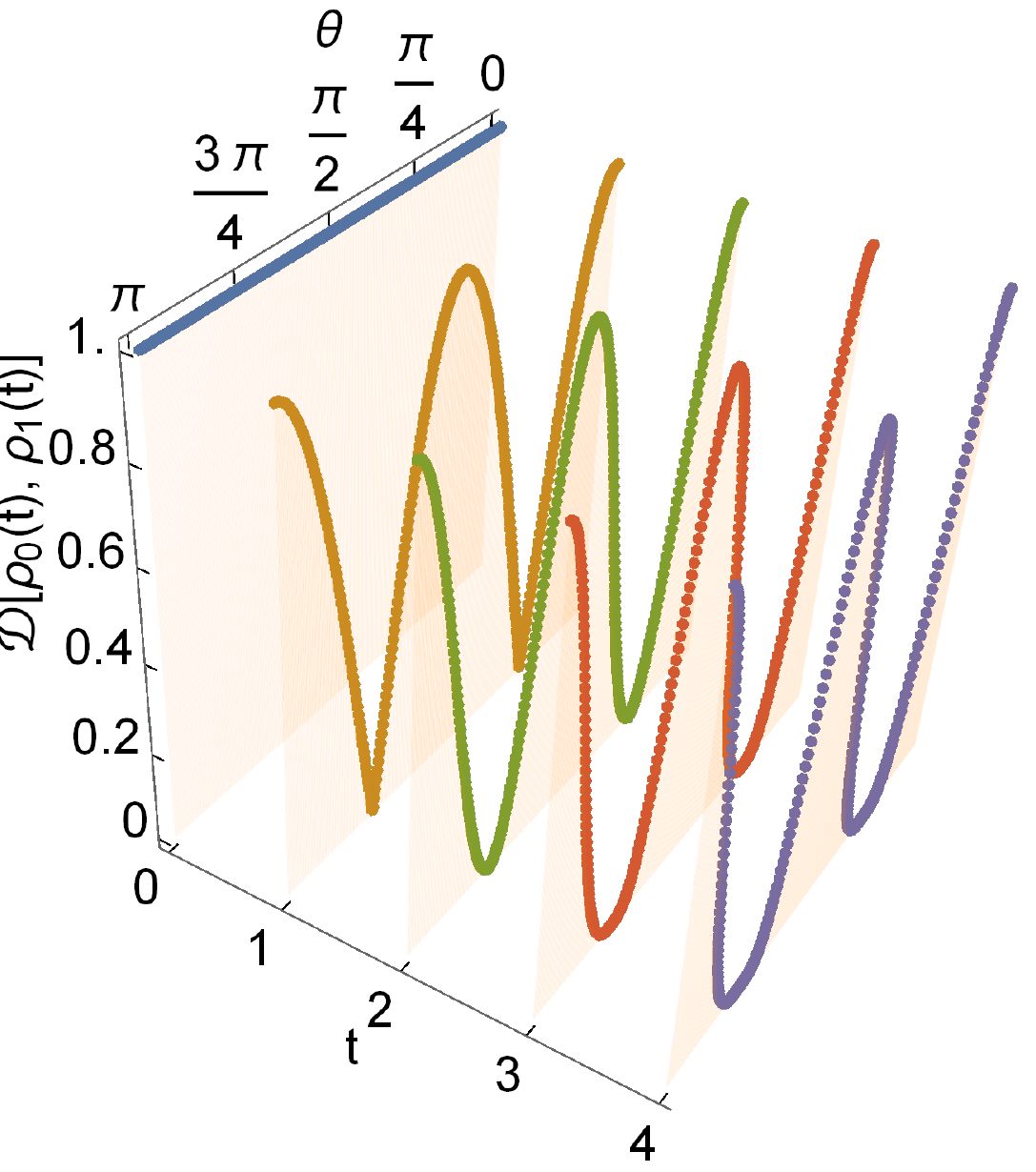} & 	\includegraphics[width=60mm,height=50mm]{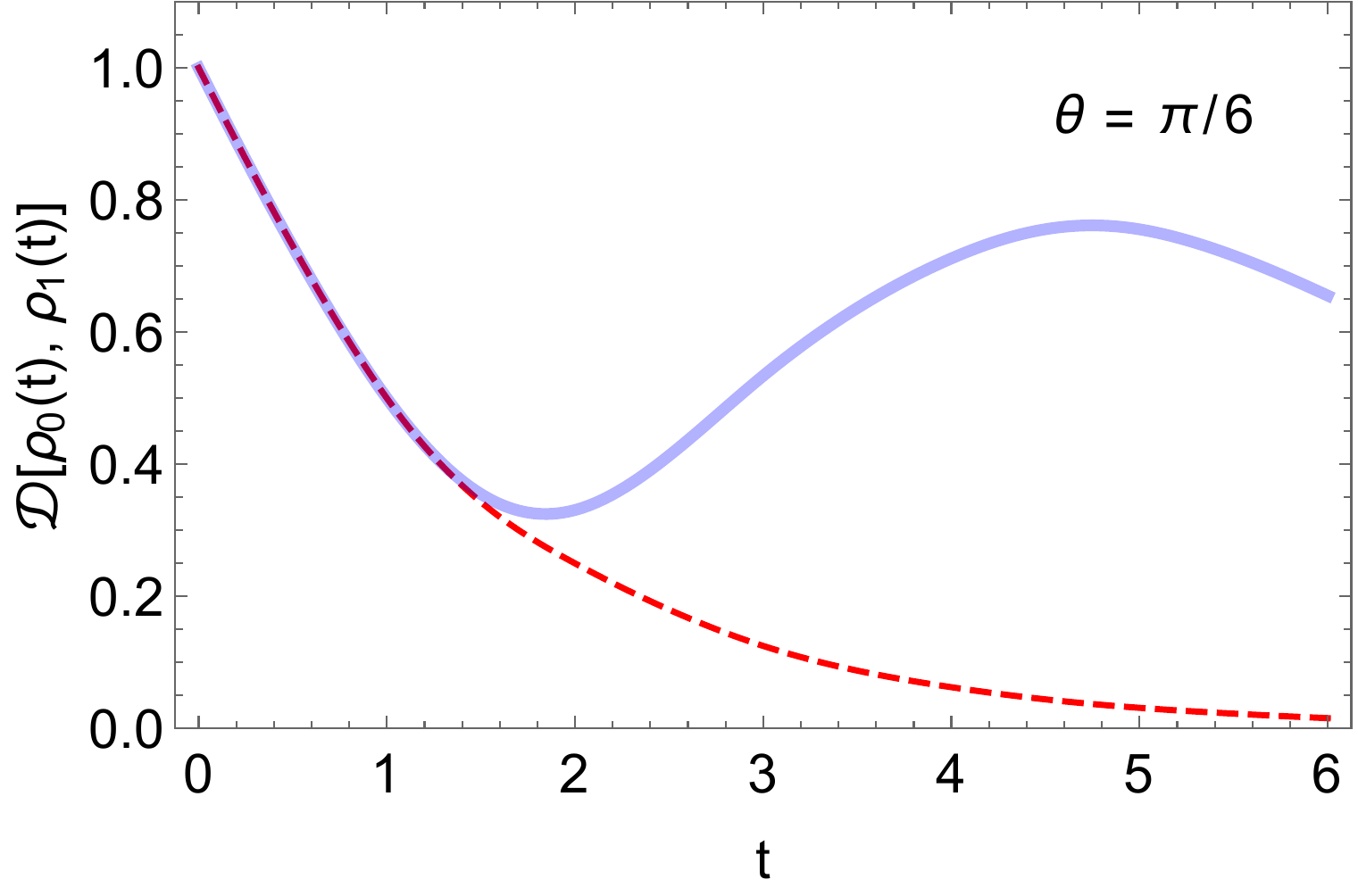}
			\tabularnewline
			(a)  & (b) & (c) \tabularnewline
		\end{tabular}
		\caption{(Color online) (a) Trace-distance between orthogonal states $\ket{0}$ and $\ket{1}$ subjected to coin dynamics as a function of coin parameter $\theta$ and the number of steps. The $n$th step is realized by applying $\mathcal{F}_n$, defined in Eq. (\ref{eq:coinmap}). (b) Trace distance between  $\ket{0}$ and $\ket{1}$ obtained by subjecting them to an  $n$-concatenation of $\mathcal{F}_1$. In  (c), we  compare (a) and (b) for $\theta = \pi/6$, with blud (solid) and red (dashed) curves corresponding to single $n$-step operation, and an $n$-concatenation operation, respectively.}
		\label{fig:TD}
	\end{figure}

	Similarly, we can compute the trace distance between $\rho_0(n)$ and $ \rho_1(n)$ for arbitrary $n$-number of steps, and is depicted in  Fig. \ref{fig:TD} (a).  The fluctuating nature of the curves clearly bring out the P-indivisibility of the non-Markovian QW channel comprising the reduced coin dynamics.
		\begin{figure}
		\centering
		\includegraphics[width=60mm]{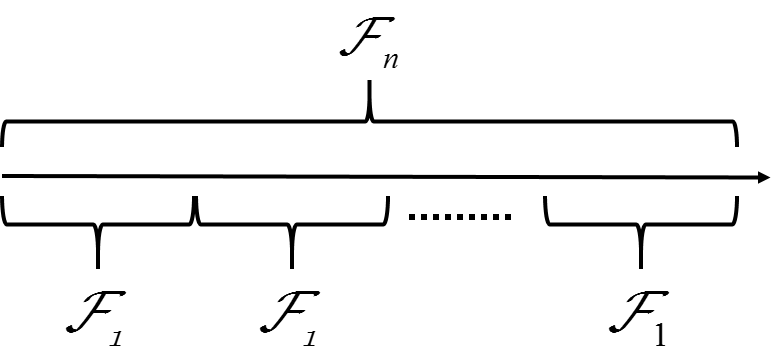}
		\caption{ The $n$-step reduced coin operation obtained by two inequivalent ways. The map $\mathcal{F}_n$ is defined in Eq. (\ref{eq:coinmap}). }
		\label{fig:nStep}
	\end{figure}

	It is important to highlight the fact that for the case of non-Markovian processes, such as the P-indivisible case studied here, the concatenation of one step map $n$ times is not equivalent to operating with $n$-step map, that is, $	\mathcal{F}_1 \mathcal{F}_1 \cdots  \mathcal{F}_1  \neq \mathcal{F}_n$, Fig. (\ref{fig:nStep}).  This becomes clear when one computes the trace distance between $\ket{0}$ and $\ket{1}$, which turns out to be a monotonically decreasing function in the former case
	\begin{equation}
     \mathcal{D} \Big[ \Big( \mathcal{F}_1   \mathcal{F}_1 \cdots  \mathcal{F}_1 \Big) \rho_0,   \Big( \mathcal{F}_1   \mathcal{F}_1 \cdots   \mathcal{F}_1 \Big) \rho_1 \Big]  = |\cos(2 \theta)|^n.
	\end{equation}
	Unless $2 \theta =0,~\pi,~2\pi$,  we have $0\le |\cos (2\theta)| < 1$, therefore, $ |\cos (2\theta)|^n $ converges to zero as $n$ increases, as shown in Fig. \ref{fig:TD} (c).
\bigskip

\begin{figure}
	      	\includegraphics[width=100mm]{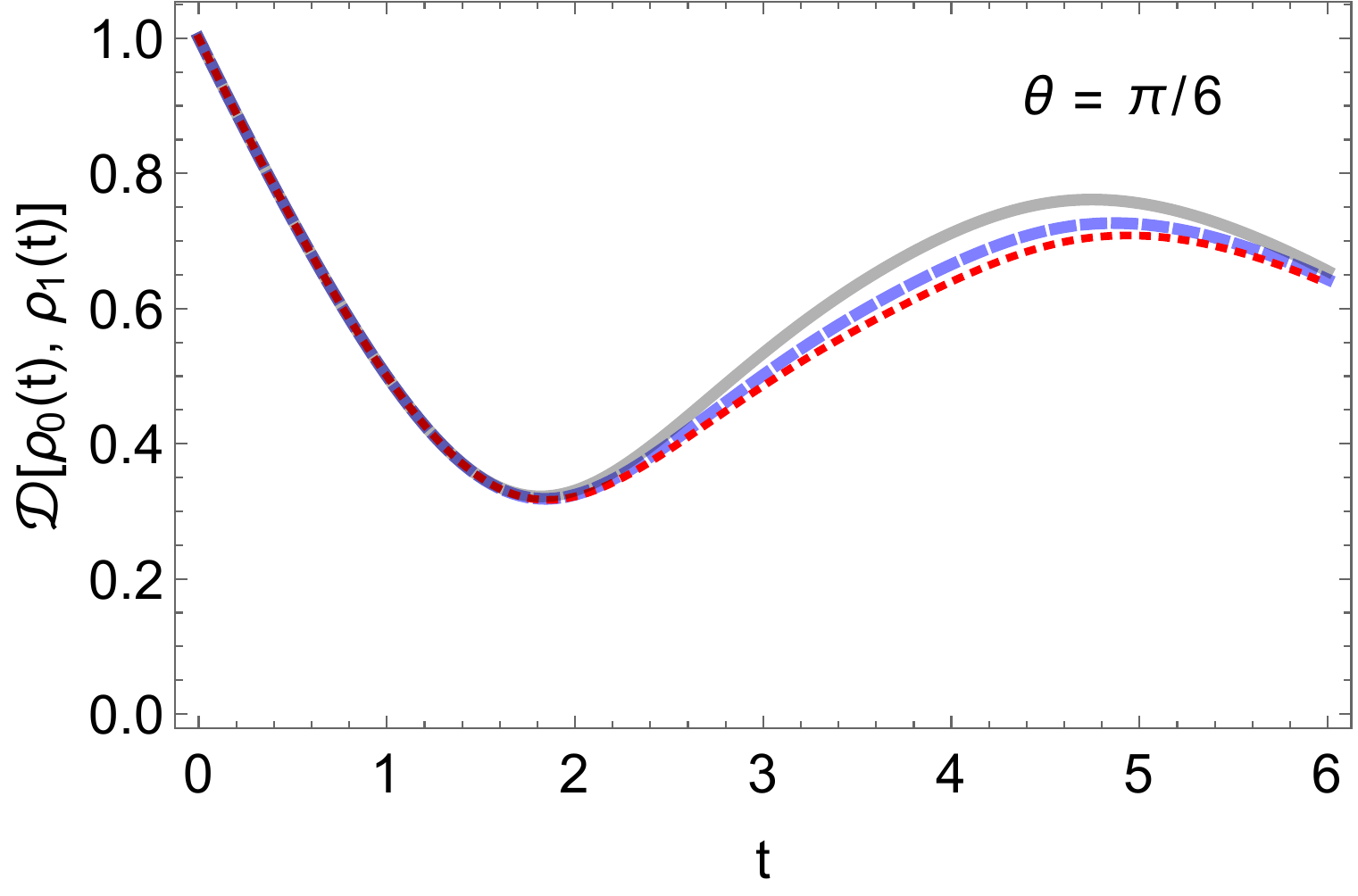}
	      	\caption{(Color online) Trace-distance between  states $\mathcal{E} \mathcal{F} ( |0\rangle \langle 0| )$ and $\mathcal{E} \mathcal{F} ( |1 \rangle \langle 1| )$, where the  composite map $\mathcal{E} \mathcal{F}$ is defined in Eq. (\ref{eq:compositeMap}). The blue (dashed) and red (dotted) curves  correspond to the cases when RTN is operated in  Markovian and non-Markovian regimes, respectively. The black curve depict the case in absence of RTN channel. The unexpected  non-monotonous behavior of trace distance in the Markovian regime  of RTN is due to the inherent non-Markovian nature of the dynamics.}
	      	\label{fig:TDRTN}
	      \end{figure}

	 \textit{Discerning multiple non-Markovian effects}: Quantum walks have been studied in the presence of various noise models, both Markovian and non-Markovian \cite{kumar2018enhanced,PradeepOSID}. It is  important to mention here that the inferences drawn about the non-Markovian behavior in such cases must take into account the inherent  non-Markovian  nature of the reduced coin dynamics. To illustrate this point, let us subject the reduced coin state to the random telegraph noise (RTN) channel, $\mathcal{E}: \rho(t) = \mathcal{E} \rho(0)$, described by following Kraus operators
	      \begin{equation}
	      R_1 = \sqrt{\frac{1 + \Lambda(t)}{2}} \mathbb{1},\qquad R_2 =  \sqrt{\frac{1 - \Lambda(t)}{2}} \sigma_z.
	      \end{equation}
	      Here,
	      \begin{equation}
	      \Lambda(t) = e^{-\gamma t} \Big[\cos \bigg(  \gamma t \sqrt{4 \frac{a^2}{\gamma^2} - 1}~\bigg) + \frac{1}{\sqrt{4 \frac{a^2}{\gamma^2} - 1}} \sin \bigg( \gamma t \sqrt{4 \frac{a^2}{\gamma^2} - 1}~\bigg)\Big].
	      \end{equation}
	  The channel describes a Markovian (non-Markovian) evolution if $\frac{a^2}{\gamma^2} < 0.25$ ( $\frac{a^2}{\gamma^2} > 0.25$). Next, we define the composition of RTN and QW channels as  $[\mathcal{E} \mathcal{F}]_{t=n}$ for n-steps, such that  
	      \begin{equation}\label{eq:compositeMap}
	        \rho_c(t=n) = [\mathcal{E} \mathcal{F}]_{t=n} \rho_c(0) = [\mathcal{E} [ \mathcal{F} \rho_c(0)]]_{t=n},
	      \end{equation}
	      where the map $\mathcal{F}$ is defined in Eq. (\ref{eq:coinmap}). Figure (\ref{fig:TDRTN}) depicts the behavior of trace distance under this composite map,  where RTN is operated both in Markovian and non-Markovian regimes. The non-monotonic behavior of trace distance in the Markovian regime of RTN channel is  a consequence of the inherent non-Makovian nature of the reduced coin dynamics.

      \subsection{Purity and mixedness under QW channel}\label{subsec:purity}
\begin{figure}
      			\centering
      		\begin{tabular}{cc}
      			\includegraphics[width=80mm]{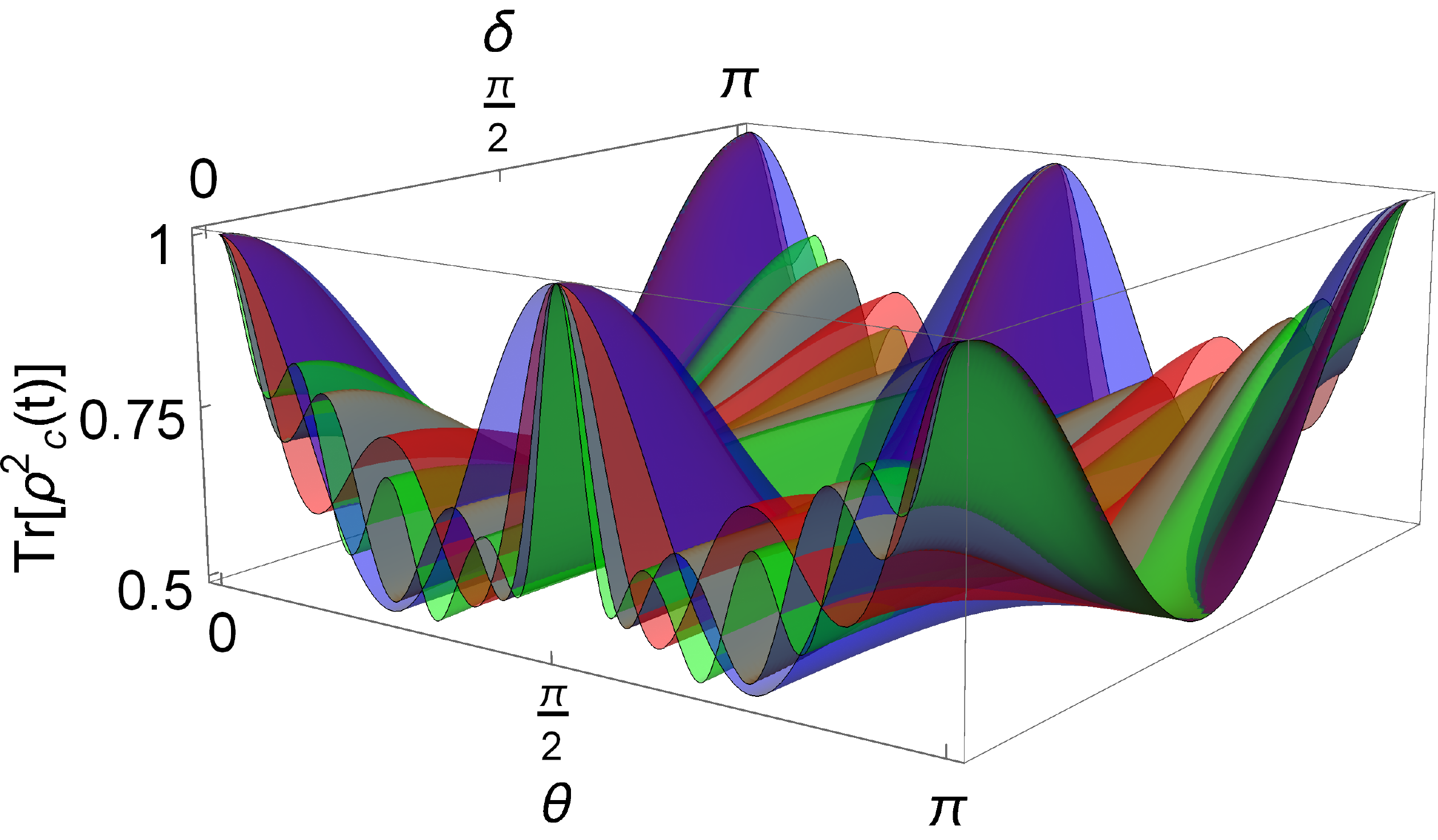} & \includegraphics[width=76mm]{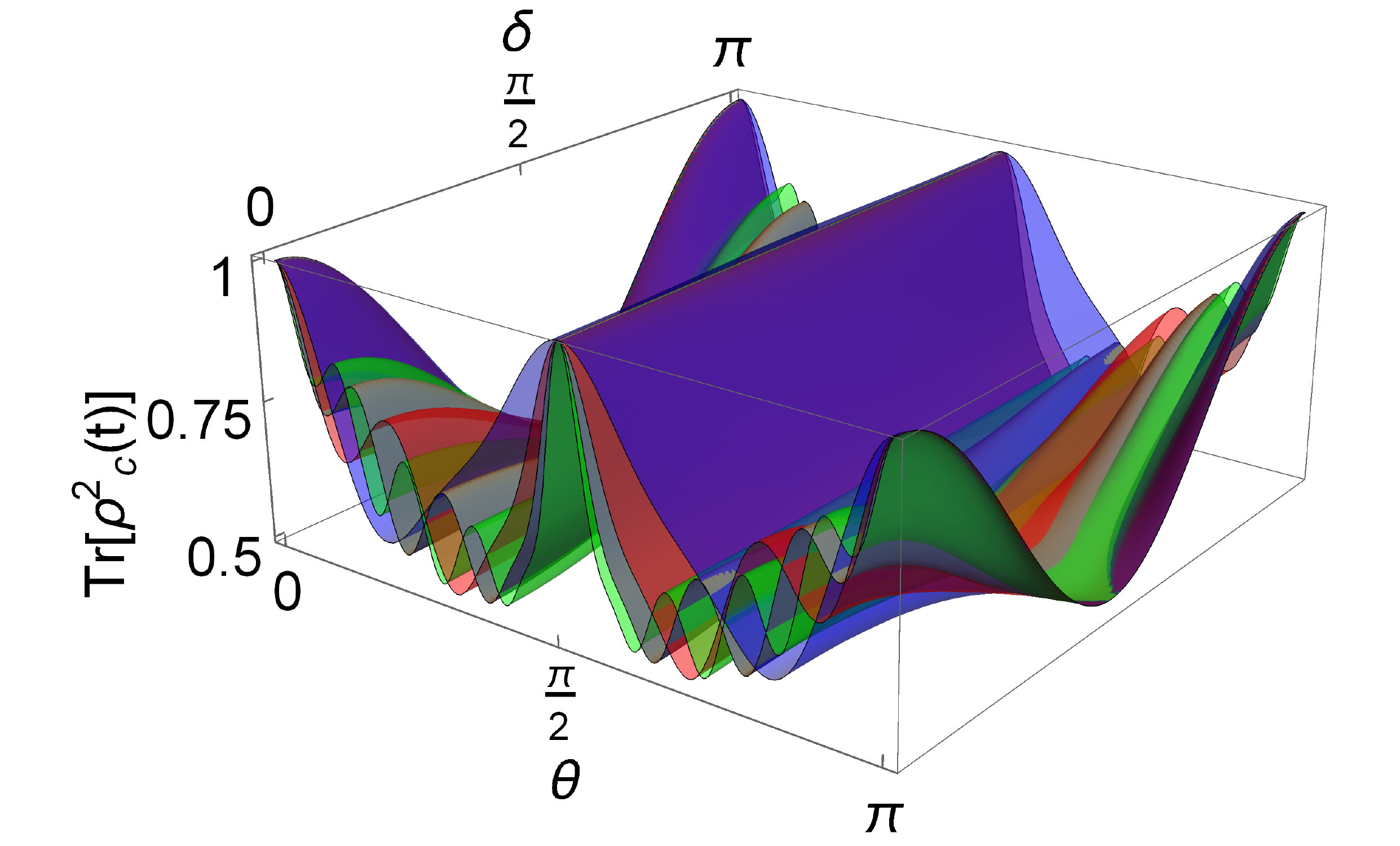}
      			\tabularnewline
      			(a)  & (b) \tabularnewline
      			\includegraphics[width=80mm]{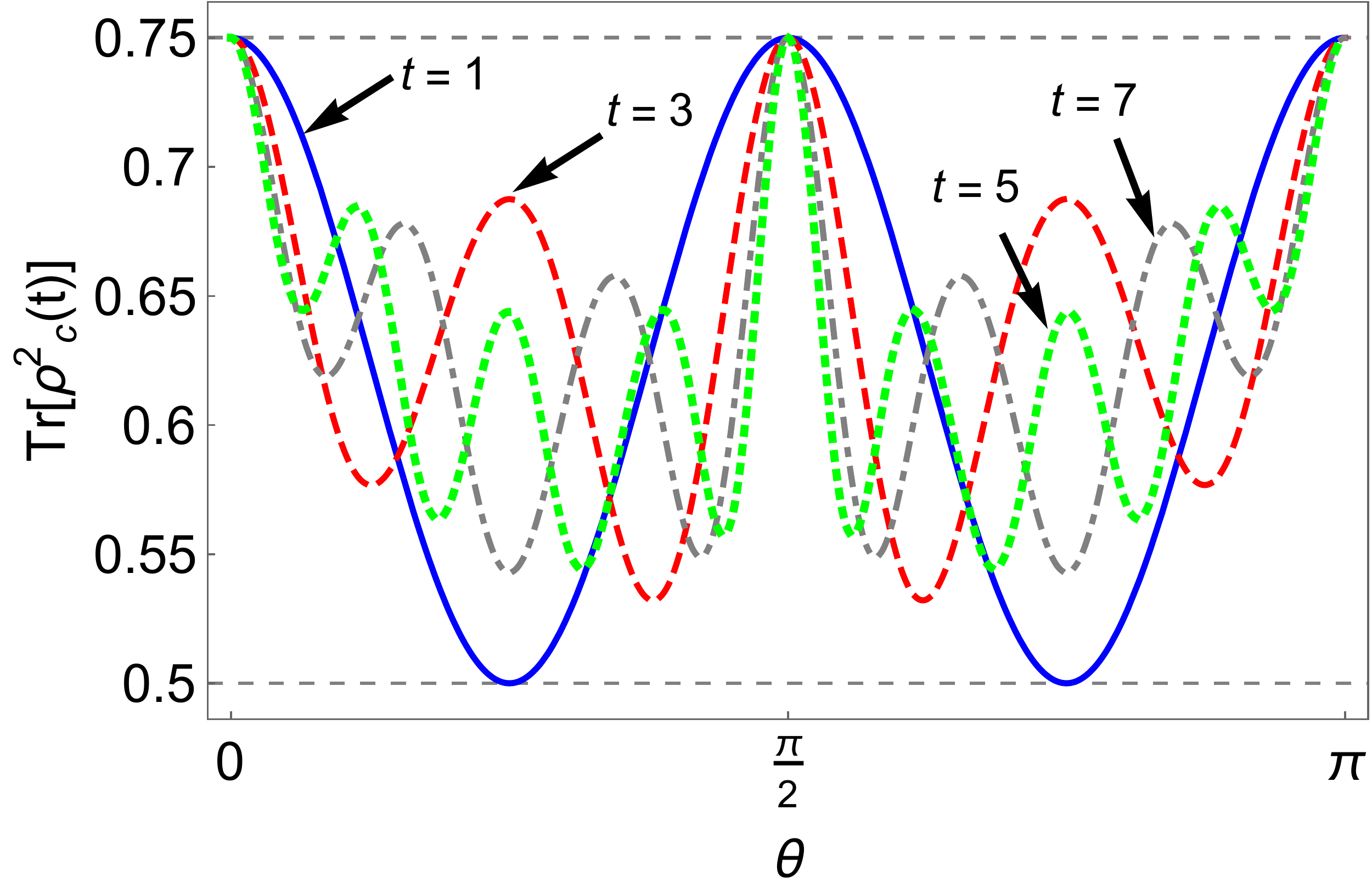} & \includegraphics[width=80mm]{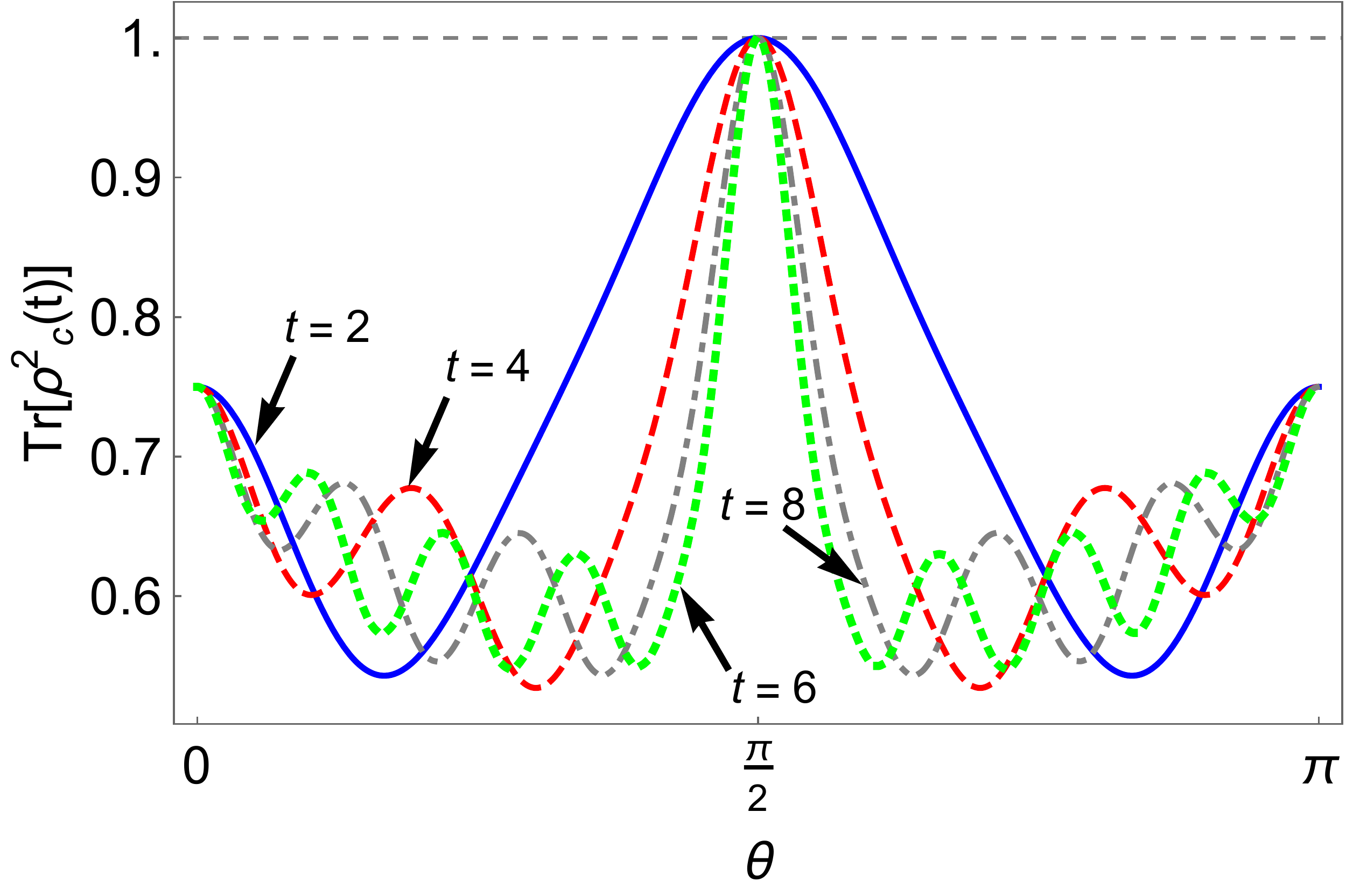}
      			\tabularnewline
      			(c)  & (d) \tabularnewline
      		\end{tabular}
      	\caption{(Color online) (a)-(b) Depicting the trace of the reduced coin state for a $t$-step QW as a function of the coin parameter $\theta$ and state parameter $\delta$ with with input state $\cos(\delta/2) \ket{0} + \sin(\delta/2) \ket{1}$. In (a) and (b) the blue, red, gray, and green surfaces correspond to $t=1,3,5,7$, and $t=2,4,6,8$,  respectively. The same quantity is plotted in  (c)-(d) with respect to $\theta$, and $\delta=\pi/4$.}
      	\label{fig:purity}
      \end{figure}

                   The purity of a state quantifies the degree of disorder or  mixedness in it. The system-environment interaction is often accompanied with a loss of coherence in the state leading to mixedness.  Thus,  purity and mixedness are complementary quantities connected by the following relation \cite{singh2015maximally}
                   \begin{equation}
                   \mathcal{M} = \frac{d}{d-1} \Big(1- Tr[\rho^2]\Big).
                   \end{equation}
                        Here, $\mathcal{M}$ is the mixedness and $Tr[\rho^2]$ is the purity of the $d$-dimensional state $\rho$. Figure \ref{fig:purity} (a)-(b) depict the purity of the output state of QW channel when the input state is $\cos(\delta/2) \ket{0} + \sin(\delta/2) \ket{1}$. For both even and odd number steps, the system is found to be in pure state for $\theta = 0, \pi/2, \pi$. The same quantity is depicted in  Fig. \ref{fig:purity} (c)-(d), with respect to the coin parameter  $\theta$, for state parameter $\delta = \pi/4$. 
      
      \subsection{Holevo quantity for QW channel}\label{subsec:Holevo}
\begin{figure}
      	\includegraphics[width=100mm]{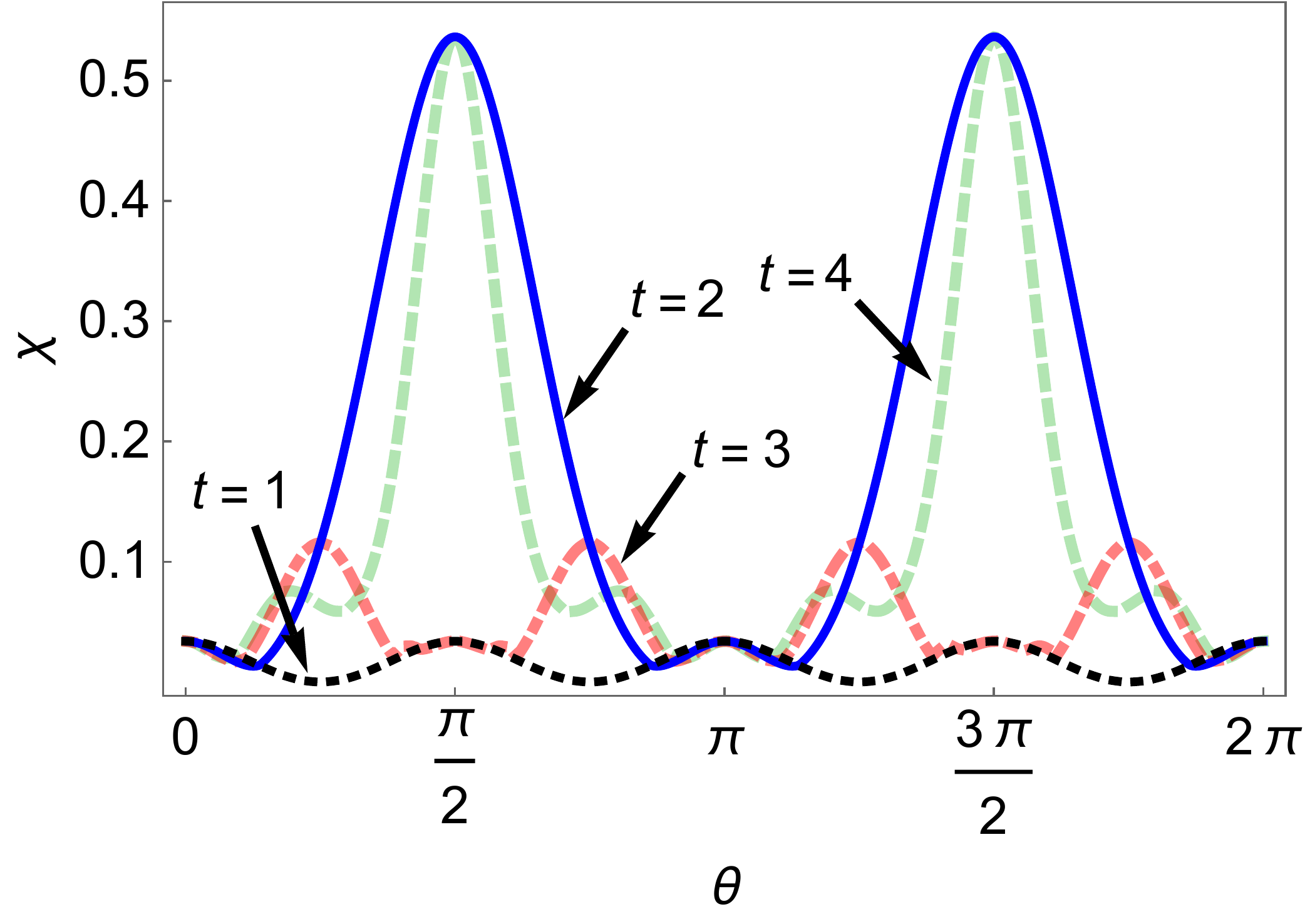}
      	\caption{(Color online)  Maximum of the Holevo quantity $\chi$ as defined in Eq. (\ref{eq:chi}). The input state is taken to be  $\rho = p_1 \rho_1 + p_2 \rho_2$, with  $\rho_1 = \frac{1}{4} | 0 \rangle \langle 0| + \frac{3}{4} |1 \rangle \langle 1|$ and $\rho_2 = \frac{1}{6} | + \rangle \langle +| + \frac{5}{3} |- \rangle \langle -|$. The maximization is carried over all $0 \le p_1 <1$ and $0 \le p_2 < 1$, constrained to $p_1 + p_2 =1$.}
      	\label{fig:Holevo}
      \end{figure}

                             When a state is subjected to a noise channel, its quantum features get affected, usually manifested in the form of decoherence and dissipation. The amount  of information about  the input state  that can be retrieved from the output state is known as \textit{accessible information}. The accessible information is upper bounded by the Holevo quantity \cite{sgad} defined as 
                             \begin{align}\label{eq:chi}
                             \chi = S\Big(\sum\limits_j p_j \mathcal{F} (\rho_j) \Big) - \sum\limits_j p_j S\Big(\mathcal{F}(\rho_j)\Big).
                             \end{align}
                             Here, $\rho_j$ is the set of input states with probability with probability $p_j$, describing the  ensemble $\{p_j, \rho_j\}$. The map $\mathcal{F}$ in our case, represents the reduced coin dynamics, and  is defined in Eq. (\ref{eq:coinmap}). Let us consider a case when the input state is described by the ensemble  $\{p_1 \rho_1, p_2 \rho_2\}$, with $\rho_1 = \frac{1}{4} | 0 \rangle \langle 0| + \frac{3}{4} |1 \rangle \langle 1|$ and $\rho_2 = \frac{1}{6} | + \rangle \langle +| + \frac{5}{6} |- \rangle \langle -|$.  For different number of steps, the Holevo quantity, maximized over $0 \le p_1 <1$ and $0 \le p_2 <1$, with $p_1 + p_2 =1$,  is depicted  in Fig. (\ref{fig:Holevo}). One infers that the Holevo quantity is suppressed for odd number of  steps.
      
      \section{Conclusion}\label{sec:conclusion}
      Recent studies have reported the constructive role of non-Markovian quantum channels over Markovian ones, in enhancing various  quantum features of the system.
      We have characterized the  reduced coin dynamics in DTQW as a non-Markovian quantum channel by  analytically computing  the Kraus operators for a  $t$-step walk. The non-Markovianity is inferred  from the P-divisibility,  reflected by non-monotonous behavior of the trace distance between two orthogonal states subjected to the channel. Subtleties  arising due to concatenation of one step map for $t$ number of steps are highlighted. This could be envisaged to have impact on the study of memory processes on  QW evolutions. The impact of  noisy channel on the purity of a quantum state is studied with respect to the number of steps as well as  the  channel (coin) parameter. The amount of information about an input state which can be retrieved from the output, is bounded by Holevo quantity, and is shown to exhibit   different behavior for even and odd number of steps.  The QW channels, introduced here,  add to the important class of non-Markovian channels which  help in developing characterization methods for open quantum systems and strategies for various quantum information tasks.    Feasibility of experimental implementation of DTQW in various quantum systems can lead way towards practical realization of non-Markovian quantum channels presented in this work.

  \section*{Acknowledgment}
  \small JN would like to acknowledge the support from The Institute of Mathematical Sciences, Chennai  to visit them during the completion of this work. 
  CMC would like to acknowledge the support from DST, government of India under Ramanujan Fellowship grant no. SB/S2/RJN-192/2014.
   \normalsize

\newpage

 \section*{Appendix}
\underline{\textit{ Calculation of $  \langle x = \mu | \hat{S}_L^k   \hat{S}_R^{t-k} | x = \nu  \rangle$}}:
\bigskip

From the definition 
\begin{align}
\hat{S}_L = \sum\limits_{x = -t}^{t} | x - 1\rangle \langle x|, \qquad {\rm and } \qquad \hat{S}_R = \sum\limits_{x = -t}^{t} | x + 1\rangle \langle x|.
\end{align}

Note  that $ \sum\limits_{x = -t}^{t} | x - 1\rangle \langle x| = \sum\limits_{x = - t -1}^{t-1} | x \rangle \langle x + 1 |$. We propose 
\begin{align}
[\hat{S}_L]^k =  \Big[ \sum\limits_{x = -t-1}^{t-1} | x \rangle \langle x + 1 | \Big]^k = \sum\limits_{x = -t-1}^{t-k} | x \rangle \langle x + k|.
\end{align}
We will prove this by induction. The cases with $k=0$ and $k=1$ trivially hold. Let us assume the results holds for $k=p$, so that 
\begin{align}\label{eq:SL}
\Big[ \sum\limits_{x = -t-1}^{t-1} | x  \rangle \langle x + 1 | \Big]^{p+1} &= \Big[ \sum\limits_{x = -t-1}^{t-1} | x \rangle \langle x + 1 | \Big]  \Big[ \sum\limits_{x = - t -1}^{t-1} | x  \rangle \langle x + 1 | \Big]^p \nonumber \\
 &= \Big[ \sum\limits_{x = -t-1}^{t-1} | x  \rangle \langle x + 1 | \Big]  \Big[ \sum\limits_{y = - t -1}^{t-p} | y    \rangle \langle y + p  | \Big] \nonumber \\
 &=  \sum\limits_{x = -t-1}^{t-1}  \sum\limits_{y = - t -1}^{t-p} | x  \rangle \langle x + 1  | y    \rangle \langle y + p  | \nonumber \\
   &=  \sum\limits_{x = -t-1}^{t-1}  \sum\limits_{y = - t -1}^{t-1} | x  \rangle \langle y + p  | ~~\delta_{x+1, y} \nonumber \\
    &=  \sum\limits_{x = -t-p}^{t- (p+1)}   | x  \rangle \langle x + p +1  |.
\end{align}
The upper limit of $x$ is restricted to $t-(p+1)$, since $y = x +1$, therefore, for  $x > t-(p+1)$ we  have  $y > t-p $, that is greater than the original limit of $y$.  Similarly, one can show
\begin{align}\label{eq:SR}
[\hat{S}_R]^k &= \Big[\sum\limits_{x = -t}^{t} | x + 1\rangle \langle x| \Big]^t = \sum\limits_{x = -t}^{t - (k-1)} |x+k \rangle \langle x|.
\end{align}

Using Eqs. (\ref{eq:SL}) and (\ref{eq:SR}), we have 
\begin{align}
\langle x = \mu | \hat{S}_L^k   \hat{S}_R^{t-k} | x = \nu  \rangle &=  \langle x = \mu | \Big[ \sum\limits_{x = -t-1}^{t-k} | x \rangle \langle x + k| \Big]  \Big[\sum\limits_{y = -t}^{t - (k-1)} |y+t - k \rangle \langle y|\Big] | x = \nu  \rangle \nonumber \\
 &= \sum\limits_{x = -t-1}^{t-k} \sum\limits_{y = -t}^{t - (k-1)} \delta_{\mu, x}  \langle x + k| y+t - k \rangle \delta_{y, \nu} =  \langle \mu  + k| \nu +t - k \rangle.
\end{align}
Therefore, this quantity is non zero for $k = (t + \nu - \mu)/2$.
 
 	\begin{table*}[htp]
 	\centering
 	\caption{\ttfamily Kraus operators for the reduced coin dynamics for some steps in a  split step quantum walk. 
 	}
 	\begin{tabular}{ |p{1cm}|p{13cm}|}
 		\hline
 		Steps       & Kraus operators \\
 		\hline
 		1               &  \parbox{3cm}{\begin{align*}
 				K_{-1} &= \left(
 			\begin{array}{cc}
 			\cos ^2(\theta ) & -i \cos (\theta ) \sin (\theta ) \\
 			0 & 0 \\
 			\end{array}
 			\right)\\
 			K_{0} &=\left(
 			\begin{array}{cc}
 			-\sin ^2(\theta ) & -i \cos (\theta ) \sin (\theta ) \\
 			-i \cos (\theta ) \sin (\theta ) & -\sin ^2(\theta ) \\
 			\end{array}
 			\right)\\
 			K_1 &= \left(
 			\begin{array}{cc}
 			0 & 0 \\
 			-i \cos (\theta ) \sin (\theta ) & \cos ^2(\theta ) \\
 			\end{array}
 			\right)
 			\end{align*}}     \\
 		\hline
 		2               &  	\parbox{3cm}{\begin{align*}
 		                                         K_{-2} &= \left(
 		                                       \begin{array}{cc}
 		                                       -\cos ^2(\theta ) \sin ^2(\theta ) & -i \cos ^3(\theta ) \sin (\theta ) \\
 		                                       -\frac{1}{4} i (3 \cos (2 \theta )-1) \sin (2 \theta ) & -3 \cos ^2(\theta ) \sin ^2(\theta ) \\
 		                                       \end{array}
 		                                       \right)\\
 		                                       K_{-1} &= \left(
 		                                       \begin{array}{cc}
 		                                       \cos ^4(\theta ) & -i \cos ^3(\theta ) \sin (\theta ) \\
 		                                       0 & 0 \\
 		                                       \end{array}
 		                                       \right)\\
 		                                       K_0 &= \left(
 		                                       \begin{array}{cc}
 		                                       \sin ^4(\theta )-2 \cos ^2(\theta ) \sin ^2(\theta ) & -\frac{1}{4} i (3 \cos (2 \theta )-1) \sin (2 \theta ) \\
 		                                       -\frac{1}{4} i (3 \cos (2 \theta )-1) \sin (2 \theta ) & \sin ^4(\theta )-2 \cos ^2(\theta ) \sin ^2(\theta ) \\
 		                                       \end{array}
 		                                       \right)\\
 		                                       K_1 &= \left(
 		                                       \begin{array}{cc}
 		                                       0 & 0 \\
 		                                       -i \cos ^3(\theta ) \sin (\theta ) & \cos ^4(\theta ) \\
 		                                       \end{array}
 		                                       \right)\\
 		                                       K_2 &= \left(
 		                                       \begin{array}{cc}
 		                                       -3 \cos ^2(\theta ) \sin ^2(\theta ) & -\frac{1}{4} i (3 \cos (2 \theta )-1) \sin (2 \theta ) \\
 		                                       -i \cos ^3(\theta ) \sin (\theta ) & -\cos ^2(\theta ) \sin ^2(\theta ) \\
 		                                       \end{array}
 		                                       \right)
 		 	                              \end{align*}}                                                                                                                                \\
 		\hline
 		3               &    	\parbox{3cm}{\begin{align*}
 		                                                    K_{-3}&= \left(
 		                                                  \begin{array}{cc}
 		                                                  -5 \cos ^4(\theta ) \sin ^2(\theta ) & -\frac{1}{2} i \cos ^3(\theta ) (5 \cos (2 \theta )-3) \sin (\theta ) \\
 		                                                  -i \cos ^5(\theta ) \sin (\theta ) & -\cos ^4(\theta ) \sin ^2(\theta ) \\
 		                                                  \end{array}
 		                                                  \right)\\
 		                                                  K_{-2} &= \left(
 		                                                  \begin{array}{cc}
 		                                                  \frac{1}{8} (1-5 \cos (2 \theta )) \sin ^2(2 \theta ) & -\frac{1}{2} i \cos ^3(\theta ) (5 \cos (2 \theta )-3) \sin (\theta ) \\
 		                                                  -\frac{1}{16} i (\sin (2 \theta )-4 \sin (4 \theta )+5 \sin (6 \theta )) & \frac{1}{4} (1-5 \cos (2 \theta )) \sin ^2(2 \theta ) \\
 		                                                  \end{array}
 		                                                  \right)\\
 		                                                  K_{-1} &=\left(
 		                                                  \begin{array}{cc}
 		                                                  \cos ^6(\theta ) & -i \cos ^5(\theta ) \sin (\theta ) \\
 		                                                  0 & 0 \\
 		                                                  \end{array}
 		                                                  \right)\\
 		                                                  K_0 &= \left(
 		                                                  \begin{array}{cc}
 		                                                  -\frac{1}{4} (4 \cos (2 \theta )+5 \cos (4 \theta )+3) \sin ^2(\theta ) & -\frac{1}{16} i (\sin (2 \theta )-4 \sin (4 \theta )+5 \sin (6 \theta )) \\
 		                                                  -\frac{1}{16} i (\sin (2 \theta )-4 \sin (4 \theta )+5 \sin (6 \theta )) & -\frac{1}{4} (4 \cos (2 \theta )+5 \cos (4 \theta )+3) \sin ^2(\theta ) \\
 		                                                  \end{array}
 		                                                  \right)\\
 		                                                  K_{1} &=\left(
 		                                                  \begin{array}{cc}
 		                                                  0 & 0 \\
 		                                                  -i \cos ^5(\theta ) \sin (\theta ) & \cos ^6(\theta ) \\
 		                                                  \end{array}
 		                                                  \right)\\
 		                                                  K_2 &=  \left(
 		                                                  \begin{array}{cc}
 		                                                  \frac{1}{4} (1-5 \cos (2 \theta )) \sin ^2(2 \theta ) & -\frac{1}{16} i (\sin (2 \theta )-4 \sin (4 \theta )+5 \sin (6 \theta )) \\
 		                                                  -\frac{1}{2} i \cos ^3(\theta ) (5 \cos (2 \theta )-3) \sin (\theta ) & \frac{1}{8} (1-5 \cos (2 \theta )) \sin ^2(2 \theta ) \\
 		                                                  \end{array}
 		                                                  \right)\\
 		                                                  K_3 &=  \left(
 		                                                  \begin{array}{cc}
 		                                                  -\cos ^4(\theta ) \sin ^2(\theta ) & -i \cos ^5(\theta ) \sin (\theta ) \\
 		                                                  -\frac{1}{2} i \cos ^3(\theta ) (5 \cos (2 \theta )-3) \sin (\theta ) & -5 \cos ^4(\theta ) \sin ^2(\theta ) \\
 		                                                  \end{array}
 		                                                  \right)
 			                                 \end{align*}} \\
 	\hline
 	\end{tabular}\label{tab:symSSQW}
 \end{table*}
 \FloatBarrier


%
\end{document}